\documentclass[preprint,twocolumn,secnumarabic,amssymb,notitlepage,nobibnotes, nofootinbib,10pt]{revtex4-1}
\usepackage{graphicx}
\usepackage[pdftex,bookmarks,colorlinks,breaklinks]{hyperref}  
\usepackage{bm}
\setcitestyle{numbers}
\newcommand{\ber}{\begin{eqnarray}}
\newcommand{\ear}{\end{eqnarray}}
\newcommand{\bc}{\begin{center}}
\newcommand{\ec}{\end{center}}

\newcommand{\be}{\begin{equation}}
\newcommand{\ee}{\end{equation}}
\newcommand{\ba}{\begin{eqnarray}}
\newcommand{\ea}{\end{eqnarray}}
\newcommand{\bs}{\begin{subequations}}
\newcommand{\es}{\end{subequations}}

\newcommand{\forget}[1]{\iffalse#1\fi}
\newcommand{\forgetmenot}[1]{\iftrue#1\fi}

\begin{document}
\title{Multi-Fluid Theory and Cosmology: A Convective Variational Approach to Interacting Dark-Sector}
\author
{Bob Osano$^{1,2}$ and Timothy Oreta$^{1}$ \\
\small{$^{1}$Cosmology and Gravity Group, Department of Mathematics and Applied Mathematics,\\ University of Cape Town (UCT), Rondebosch 7701, Cape Town, South Africa\\}
\small{$^{2}$Centre for Higher Education Development,\\ University of Cape Town (UCT), Rondebosch 7701,Cape Town, South Africa}\\
}



\begin{abstract} This article examines the foundation of the recently developed {\it relativistic variational formalism}\cite{Cart0}. Our work is heavily based on \cite{And5, Comer2} which extends this approach to the multi-fluid theory and examines its utility in astrophysics and cosmology. Unlike the extension to the formalism mentioned above, that looks at the general interaction between different types of matter, we use the formalism to examine the interaction involving, ordinary matter, dark matter ($DM$) and dark energy ($DE$). We focus on an {\it entrainment} phenomena involving the dark-sector constituents.
\nopagebreak
\end{abstract}
\date{\today}
\maketitle
\section[1]{\label{sec1} Introduction}
The theory of relativistic fluids has received considerable attention for many years following the seminal work of $Landau$ and $Liftshitz$\cite{Land}. Interest in relativistic fluids is largely driven by its utility in astrophysics and cosmology and the need to resolve some of the outstanding questions in these areas of study. The mathematical modelling process that is carried out in these studies, and which take into account physically plausible scenarios, presents several challenges. These fall into two broad categories; {\it conceptual} and {\it theoretical}. An awareness of the differences between these two categories is important as it provides a guide to a modeller when making decisions on assumptions that go into the modelling process, and when analysing the model. It is therefore, worth reviewing the differences between the two categories. 

The conceptual related issues have to do with the difficulty in identifying specific and measurable variables that give rise to a framework for characterising relativistic fluids. Issues classified under theoretical category have to do with the foundational theories, which in this study are theories of fluids, general relativity ($GR$), and thermodynamics as applied to environments where there are more than one species of fluids. Connected to this is what might be termed as the unifying framework of how the species are treated; either single-fluid or multi-fluid. For example, the nature of interactions between the different species may affect how the mixture flows, effects that may only be captured in the multi-fluid treatment and not in the single-fluid theory. Examples of these are {\it dissipation} and {\it entrainment}. Dissipative effects are known to be common and are found in flows involving heat flow in the presence of thermal resistance, in fluid flows with viscosity, diffusion, chemical reactions, and electric current flow in resistive media. These examples and others manifest in the lab environment, in astrophysics and predictably in cosmology. Dissipation has successfully been incorporated and examined in the modelling of $Newtonian$ or non-relativistic fluids. But the same cannot be said of relativistic fluids (as pointed out by \cite{Hiscock, And5, Leo} for example). The obvious question is, what motivates the need to incorporate dissipation in relativistic fluids and how would one go about doing this? The authors of \cite{And1} are prompted by the need to develop a formalism that could be used to study gravitational radiations emanating from compact objects; neutron stars in particular. Radiative processes in some of these astrophysical objects are known to be influenced by dissipation. Dissipation is often largely neglected in cosmology \cite{Weinberg, Patel, Velten}, but there are processes that occur during structure formation, and during reheating epochs in the early universe that suggest that dissipation may play a role and hence should ideally be taken into account. The same can be said of heat flow in general \cite{Modak, Triginer, And6} and $DM$ dynamics \cite{Sikivie, Harko}. In order to account for these, one needs to develop a formalism that incorporates them. Entrainment is understood to be the quantification of the ease with which neighbouring fluids species are able to move relative to each other. Unlike dissipation, entrainment is much less known or studied particularly for the subclass of relativistic fluids.

We are motivated by the need to examine multi-fluid and entrainment effects involving the $DM$ and $DE$. As pointed out in \cite{And4}, the most interesting development in classical relativistic fluids dynamics is the consideration of multi-fluid systems that is composed of elements whose collective dynamics involve a superfluid, heat flow or the treatment of electromagnetic charge as a dynamical variable \cite{Lop1, And5, List1}. These developments are allowing such systems to be used to study a wider range of relevant phenomena. These developments have, however, been patchy and a general theory remains incomplete in at least two different respects. On one hand, they require the inclusion of dissipation and on the other, the coupling of dissipation to electromagnetism. These developments hold the key to the greater applicability of the fluid theory in both astrophysics and cosmology and need examination. Single-fluid approximation has successfully played a crucial role in our cosmological modelling of the universe. It is our contention that multi-fluid approximation is the more appropriate for cosmological modelling. The single-fluid approximation is ideally the limit of multi-fluid approximation. In this regard, the success of the single-fluid approximation in cosmological modelling is tied to the fact that different species making up the cosmological fluid may start off evolving differently but eventually, all species become {\it locked-in} thereby rendering one species dominant at a given epoch. This defines the fundamental observer world-line. The mechanism that allows this to happen is yet to be developed. The above statements, simple as they are, demand a re-examination of relativistic multi-fluid theory in particular and its application to cosmology in general. We will assume a generic model of $DE$ that interacts with $DM$ for illustrative purposes in our study. The existence of interactions between $DM$ and $DE$ has been a subject of much inquiry for the reason that they can fit well the observational data and could potentially provide new physics. For example, they might help resolve the {\it coincidence} problem. i.e. provide a possible explanation as to why the present values of the densities of $DE$ and $DM$ are of the same order of magnitude, something that would require very special initial conditions in the early universe. Unlike the well behaved non-interacting models with constant $w$ (given a barotropic equation state, where $w$ is the proportionality parameter), an interacting model can manifest instabilities in the perturbations of the dark-sector at early times giving rise to new phenomena. Interacting dark-sector models have, for example, been studied in \cite{Eliz,Far, He, Yang, Bruck, Val, Manos, Manos2}

This article is arranged as follows; section(\ref{sec2}) discusses multi-fluid and the {\it convective variational} formalism. Section (\ref{matter}) discusses $DE$ and $DM$, while section (\ref{lagr0}) discusses the $Lagrangian$ formulation involving interacting dark-sector and coresponding equations of motion. Section (\ref{sec5}) gives the discussion for the results, conclusion and future work.

\section{\label{sec2} Fluid Theory and Relativistic Convective Variational Formalism}
Our presentation in this section borrows heavily from and builds on, \cite{Cart1, Comer2, And1, And2, And4}. It is our intention to examine the dynamics in a multi-fluid environment involving fluid species that will be relevant to cosmology. In this regard, it is important first to clearly distinguish between single-fluid and multi-fluid theories.

The term {\it multi-fluid } is generally understood to refer to a mixture of fluids that is made up of many-species where each species is treated as a separate and unique fluid entity but at the same time allowed to contribute to the dynamics of the mixture. The separate treatment of species in the multi-fluid theory may, for example, allow the different species to have different temperatures thereby allowing for heat flow. In order to capture the ensuring artefacts manifested in the treatment, the system of equations in the multi-fluid approach could include equations of energy for the individual species in comparison to a single energy equation for the mixture used in the single-fluid approach. Mean velocities of individual species are specified by their respective momentum equations \cite{Beni} in multi-fluid approach. The system of fluid equations include both momentum equations for the whole mixture and the transport equations for the individual species. In this approach, it is imperative that the momentum equations include the convective terms, which are normally absent in the transport equations. The multi-fluid approach, therefore, suggests the existence of several fundamental observers, each with its own 4-velocity. Multiple velocities suggest an anisotropic model, for example, a $Bianchi$ type I model \cite{Tsagas, Bob1}. Current observations indicate that our universe is isotropic and homogenous on large scales. The question whether it started out this way or transitioned into this state is still being investigated. It is, therefore, useful to require that any anisotropic model, such as those resulting from a multi-fluid approximation, must become isotropic if they are to be taken as an alternative to the standard model. It has been shown \cite{Comer2, Harko2} that indeed isotropization occurs in a two fluid-model, suggesting a possible extension to a general multi-fluid model. We will require isotropization as a condition for the multi-fluid setup in this article. In fact, the authors of \cite{Comer2} have found that there exists a $Bianchi$ type I epoch where the matter flux dominates which eventually evolves to $Friedmann-Lemaitre-Robertson-Walker$ $FLRW$ model; effectively a single-fluid model. We will demonstrate, in an accompanying article \cite{Bob10}, that the second law of thermodynamics is satisfied in a three-fluid system that allows chemical reactions to isotropize in late time.
\subsection{\label{thermo1} Single-Fluid Approximation and Thermodynamics}
The author of \cite{Ted} has considered a novel way of deriving the classical $Einstein$ equations from thermodynamical considerations. In particular, he uses the heat relation $\delta Q=TdS$, the proportionality of entropy and the horizon area to arrive at $Einstein$ equations and points out the analogy between this equation and the equation of state ($EoS$), subject to local equilibrium conditions as defined by the relationships between thermodynamic variables. It is important to mention that there exists a length scale in this approximation for which the conditions are assumed to hold with no threat of the emergence of transient thermodynamics \cite{Isra1}. The causal horizon is connected to the entropy and holds information \cite{ber1} that could potentially be decoded but the tools for doing this are yet to be developed. We will not recount the full procedure for deriving the $Einstein$ equations from thermodynamics in this article but will reserve the complementary development for multi-fluid for an upcoming article \cite{Bob10}. Nevertheless, one can easily show the connection between entropy and the $EoS$. As an example, let us assume that the entropy function is known and that it is given in terms of total internal energy $E$, volume $V$, number density $N$, temperature $T$, pressure $p$ and chemical potential $\mu$. It follows from the first law of thermodynamics that $\delta Q= dE+pdV-\mu dN$ from which one can conclude that $\partial S/\partial E \equiv T^{-1}$,~~$ T \partial S/\partial N \equiv \mu$,~~$T\partial S/\partial E \equiv p$. The last relationship is just the $EoS$. In \cite{Ted} heat is defined as energy flux across a casual horizon that can be felt via the gravitational field it generates, while $T$ is the $Unruh$ temperature i.e. as measured by a uniformly accelerated observer. It is known that acceleration diverges as the observer world-line approaches the horizon, nevertheless, there exists a limit where the ratio of $Unruh$ temperature and energy flux both remain finite. It is in this limit that thermodynamics is examined in \cite{Ted}. We see or can show that their analysis is done for a single observer world-line which is akin to studying a single-fluid formulation. As previously mentioned, it is of interest to us to show how one would construct a complementary argument for multiple observer word-lines. An attempt at examining thermodynamics in multi-fluid theory is provided in \cite{Mub, Bob10} where it is shown that entropy always increases (i.e. $\dot{S}\ge 0$). Any success in this will provide a far greater understanding of thermodynamics and gravity in a multi-fluid environment.

Relativistic convective variational formalism \cite{Cart0} and its variant; flux-conservative formalism \cite{And5, And2} suggests how we might proceed. We first look at energy considerations for a single-fluid setup subject to the first and second laws of thermodynamics. Let us first examine this in the $Newtonian$ fluid. The extension to relativistic fluids will be straight forward. 
\begin{eqnarray}
dE&=&\frac{d{E}}{d{N_{n}}}dN_{n}+\frac{d{E}}{d{V}}dV+\frac{d{E}}{d{S}}dS\nonumber\\
&\equiv&\mu dN_{n}-pdV+TdS,
\end{eqnarray}
where again $E$ is the total internal energy, $N$ is the number of particles, $V$ is the volume, $S$ is the entropy, $\mu$ is the chemical potential, $p$ is the pressure and $T$ is the temperature. As in \cite{And2}, we will assume that the doubling of $E$ is a direct result of the doubling of $N$, $S$ and $V$, (i.e. extensive parameters) whereas $T$, $P$ and $\mu$ are unchanged (i.e. intensive parameters). It is easy to show that this leads to $Euler$ relation $E=TS-pV+\mu N$. Given the $Gibbs$-$Duhem$ requirement or equation, it follows that
\begin{equation}
E=\mu N-pV+TS,
\end{equation} 
implying that
\begin{eqnarray}\label{Eulers}
\frac{E}{V}&=&\mu \frac{N}{V}-p+T\frac{S}{V},
\end{eqnarray}
which can be written in terms of $\Lambda=-E/V$, $n_{n}=N/V$, $n_{S}=S/V$, $\mu=\mu^{n}$ and $T=\mu^{s}$ and which gives the relation 
\begin{eqnarray}\label{Lam}
-\Lambda+p&=&\mu^{n} n_{n}+\mu^{s}n_{s},
\end{eqnarray} 
where $\Lambda$ is the total energy density and will be central to the $Lagrangian$ formulation that we will use. Furthermore, it can be demonstrated that in co-moving parameters, this master function takes the form
\begin{eqnarray}\label{Lam2}
d\Lambda=\frac{d{\Lambda}}{d{n^{2}_{n}}}dn^{2}_{n}+\frac{d{\Lambda}}{d{n^{2}_{s}}}dn^{2}_{s},\nonumber\\
\end{eqnarray} 
where $\Lambda=\Lambda (n^{2}_{n}, n^{2}_{S})$. In essence this depicts single-fluid approximation although the master function has two components which are related to the same fluid species. It is vital to point out the context of multi-fluids vs single-fluids that is based on fluid species and not components of the same species. For example, two separate species $X$, and $Y$ would have $\Lambda_{X}=\Lambda (n^{2}_{n_{(X)}}, n^{2}_{S_{(X)}})$ and $\Lambda_{Y}=\Lambda (n^{2}_{n_{(Y)}}, n^{2}_{S_{(Y)}})$ respectively. It is also straight forward to show from equation (\ref{Lam2}) that 
\begin{eqnarray}
\frac{d{\Lambda}}{d{n^{2}_{\alpha}}}&=&-\frac{d{(\mu^{\alpha}\sqrt{n^{2}_{\alpha}})}}{d{n^{2}_{\alpha}}}=-\frac{\mu^{\alpha}}{2n_{\alpha}},
\end{eqnarray} 
where $\alpha=n, S$. This suggests that 
\begin{eqnarray}
d\Lambda=-\frac{\mu^{n}}{2n_{n}}dn^{2}_{n}-\frac{\mu^{S}}{2n_{S}}dn^{2}_{S}.\nonumber\\
\end{eqnarray} 
As in \cite {And2}, one can derive the chemical potential co-vectors which are conjugate to the  number and entropy fluxes,
\begin{eqnarray}\label{beta01}
\mu^{\alpha}=(-2\frac{d{\Lambda}}{dn^{2}_{\alpha}})n_{\alpha}=\mathcal{B}^{\alpha}n_{\alpha},
\end{eqnarray} 
where again $\alpha=n, S$. Using equation (\ref{beta01}), it is possible to rewrite equation (\ref{Lam}) in terms of co-moving coordinates only. This yields
\begin{eqnarray}
f(n^{2}_{\alpha})=\Lambda+\mathcal{B}^{\alpha}n^{2}_{\alpha}.
\end{eqnarray} 
Here too  $\alpha=n, S$. It is clear that $f(n^{2}_{\alpha})$ represents pressure which is a function of energy densities and hence gives the effective $EoS$ that is barotropic \cite{Com1}. The case for multi-fluid is not as straight forward.

\subsection{\label{thermo2} Multi-Fluid Approximation and Thermodynamics}
In this case, we take the same starting point as in the previous section but now consider two-fluid species rather than one. In particular, we have species $X$, and $Y$ which are mixed. It should be clear that we no longer have the individual $\Lambda_{X}=\Lambda (n^{2}_{n_{(X)}}, n^{2}_{S_{(X)}})$ and $\Lambda_{Y}=\Lambda (n^{2}_{n_{(Y)}}, n^{2}_{S_{(Y)}})$ respectively, but rather $\Lambda$ that now encodes contributions from both species. Following \cite{Cart0, And5}, we begin with the assumption that there exists a two-fluid environment where the different species move with individual velocities. Let $n_{(X)}$ and $n_{(Y)}$ denote the two densities and ${u}^{\mu}_{(X)}$ and $ u^{\nu}_{(Y)}$ denote the corresponding velocities, where $\mu$ and $\nu$ denote spatial directions. It follows that the fluxes for the individual species are given by $ n^{\mu}_{(X)}=n_{(X)} u^{\mu}_{(X)}$ and $n^{\nu}_{(Y)}=n_{(Y)}u^{\nu}_{(Y)}$, where we have assumed that there are no chemical interactions between the two species. Each number density is separately conserved, (i.e. $\nabla_{\mu}n^{\mu}_{X}=0=\nabla_{\mu}n^{\mu}_{Y}$). Taking the flux $n^{\mu}_{(X)}$ for each component as a fundamental field, one can derive associated co-moving densities associated with the given flux. Let $n_{(X)}^{\mu}$ and $n_{(X)}^{\nu}$ be two fluxes for fluid of type $X$ (same species) that are endowed with the spatial indices $\mu$ and $\nu$ respectively. It follows that the product $n^{\mu}_{(X)}n^{\nu}_{(X)}=n^{2}_{(X)}u^{\mu}_{(X)}u^{\nu}_{(X)}$. The co-moving density is then obtained as follows  
\begin{eqnarray}
g_{\mu\nu}n^{\mu}_{X}\overrightarrow n^{\nu}_{X}&=&g_{ab}n^{2}_{X} u^{\mu}_{X}u^{\nu}_{X}=n^{2}_{X}(g_{\mu\nu}{u}^{\mu}_{X}{u}^{\nu}_{X})\nonumber\\&=&-n_{(X)}^{2},
\end{eqnarray} 
where ${u}_{X}^{\mu}$ and $ u_{X}^{\nu}$ are the individual four velocities for the types of fluids and $g_{\mu\nu}$ is the space-time metric such that 
$g_{\mu\nu}{u}^{\mu}_{(X)}{u}^{\nu}_{(X)}={u}^{\mu}_{(X)}{u}_{\mu (X)}=-1$. This can be extended to a two-fluid model such that  $g_{\mu\nu}{n}^{\mu}_{(X)}{n}^{\nu}_{(Y)}=g_{\mu\nu}n^{2}_{(XY)}u^{\mu}_{(X)}u^{\nu}_{(Y)}=n^{2}_{(XY)}$, where $u^{\mu}_{(X)} u_{\mu(Y)}=-1$.  Taken in totality, the formulation suggests that the energy density $\Lambda$ can be expressed as a function of energy density scalars such that $\Lambda=\Lambda(n^{2}_{n_{(X)}},n^{2}_{s_{(X)}},n^{2}_{n_{(Y)}},n^{2}_{s_{(Y)}}$)
if there are no chemical interactions or $\Lambda=\Lambda(n^{2}_{n_{(X)}},n^{2}_{s_{(X)}},n^{2}_{n_{(Y)}},n^{2}_{s_{(Y)}},n^{2}_{n_{(XY)}},n^{2}_{s_{(XY)}}$) if a chemical interaction occurs leading to the momentum conjugate taking the form
\begin{eqnarray}\label{entr}
{\mu^{X}}_{\nu}&=& g_{\nu\mu}\left(\mathcal{B}^{X}{n^{\mu}}_{X}+\mathcal{A}^{XY}{n^{\mu}}_{Y}\right)\nonumber\\
{\mu^{Y}}_{\nu}&=& g_{\nu\mu}\left(\mathcal{B}^{Y}{n^{\mu}}_{X}+\mathcal{A}^{YX}{{n^{\mu}}_X}\right)
\end{eqnarray} 
where
\begin{eqnarray}
\mathcal{A}^{XY}&=&\mathcal{A}^{YX}=-\frac{\partial \Lambda}{\partial n^{2}_{XY}}, ~~X\ne Y
\end{eqnarray}
The last terms in equations (\ref{entr}) encapsulates the entrainment effect, a topic that we will return to shortly. It is important to note that this raises fundamental issues about how one might define local thermodynamics equilibrium. It suffices to say that local thermodynamic energy is recovered in the limit where all fluxes are parallel. The terms $n^{2}_{n_{(XY)}},n^{2}_{s_{(XY)}}$ are meaningful when dissipative fluids are considered, as we will do here but we first dispense with the idea of a lambda that has more than two species. Mathematically, it is easy to see that the master function could have the product $n^{4}_{n_{(WXYZ)}}=n^{2}_{n_{(WZ)}}n^{2}_{n_{(XY)}}$ as one of its entries. This is because one could potentially construct other scalar quantities by taking products of lower powered scalars i.e., $(g_{\mu\nu}n^{\mu}_{(W)}n^{\nu}_{(X)}g_{\nu\gamma}n^{\nu}_{(Y)}n^{\gamma}_{(Z)})$. Intuitively, this suggests the product of entrainment involving four species of fluids. The physical significance of such products are somewhat unclear, particularly when viewed against the notion of local thermodynamics equilibrium. We will address this in \cite{Bob10}. It is nevertheless important to list some of the couplings that may complicate the modeling process: These are (i) Matter-Matter (fluxes, flow-lines and entrainment), (ii) matter - spacetime ( fluxes and metric - stress-energy tensor), (iv) Matter - Electromagnetism (fluxes and current) and (v) Electromagnetism - Spacetime ( potential and curvature). The last two couplings are synonymous in $GR$ as spacetime curvature is at the behest of matter distribution. The first coupling is straightforward and has been mentioned above. The state of matter involved in the couplings mentioned above may, in principle, be determined thermodynamically \cite{Reichl}, where only a few parameters are monitored as the fluid changes and other associated or depended parameters are recovered via the $EoS$. This means that where the $EoS$ is known, one needs only monitor truly independent variables. But this raises the question of whether it is possible to determine or constrain the $EoS$ if the relationships between primary variables are all known. The question is not trivial given that we would like to apply the formalism to multi-fluid environment that includes $DM$ whose $EoS$ is not yet established. 

\section{\label{matter}Dark Matter and Dark Energy}
It is correct to say that the predictions of the existence of $DM$ and $DE$ come from observations. In particular, the fitting of a theoretical model to the composition of the universe given a combination of different cosmological observations leads to the predictions that the universe is made up of ~68\% $DE$, ~27\% $DM$, and ~5\% normal matter. But fitting models, apart from being predictive, does not give the physics of constituent particles. It is particularly insidious, to our pristine picture of the evolution of the universe, that the two species that we know very little about are the very species that have profound effects on the evolution of the universe resulting in structure formation in the early universe and an accelerated expansion in the late universe. Is single-fluid approximation partly to blame and could multi-fluid approximation shed any light in this? These questions touch on the very foundation of the $Copernican$ principle\cite{Chris} and the more stringent cosmological principle \cite{Labini, Hansen}. Although we are not investigating the cosmological principle in this study, the formalism suggests the need for the relaxation of the principle in some cosmological epoch.

Let us consider $DM$. Although we do not know what it is made of, we can rule out a number of candidates. These include stars, planets, baryons, anti-matter, and large galaxy-sized black holes. There are, however, a few viable $DM$ candidates. It is thought that baryonic matter tied up in brown dwarfs or heavy elements could still make up the $DM$. These possibilities are known as massive compact halo objects, or {\it MACHOs}. But the most common view is that $DM$ is not baryonic at all, but that it is made up of other, more exotic particles like axions or weakly interacting massive particles({\it WIMPS}). This means that any analysis involving $DM$ will have to make some assumptions about its nature. What we can ask at this stage is if any one of these candidates allows for entrainment. Putting it differently, can we use entrainment to distinguish characteristics between these candidates?

\section{\label{lagr0} Energy Functional: The Lagrangian}
We now develop the arguments in this section in terms of fluid action. Unlike in \cite{And5} where particles ($n$) and entropy ($s$) were treated as the two fluids in multi-fluid formalism, we have two separate but interacting fluids; $DM$ and $DE$ each having both particle and entropy components. This makes our multi-fluid to have at least 4 fluxes, which in the language of \cite{And5} is just a two-fluid system with four components. As in \cite{And5}, one can find the variation of the {\it master} functional $\Lambda$ in terms of constrained $Lagrangian$ displacement where it is varied with respect to the fluxes and the metric $g_{\mu\nu}$. It follows that one can construct the relevant $Lagrangian$; $\Lambda=\Lambda(n^{2}_{n_{(DM)}},n^{2}_{s_{(DM)}},n^{2}_{n_{(DE)}},n^{2}_{s_{(DE)}},n^{2}_{n_{(DM-DE)}},n^{2}_{s_{(DM-DE)}})$. The $Lagrangian$ variation $\Delta\equiv\delta+\mathcal{L}_{\xi}$, see appendix (\ref{pullb}), of the action involving this {\it master} function in terms of $Lagrangian$ displacement $\xi^{\nu}$ leads to,
\begin{eqnarray}
\delta(\sqrt{-g}\Lambda)&=&\frac{1}{2}\sqrt{-g}\left(\Lambda g^{\mu\nu}+g^{\lambda\nu}\sum_{X} n^{\nu}_{X}\mu^{X}_{\lambda}\right)\delta g_{\mu\nu}\nonumber\\
&-&\sqrt{-g}\sum_{X}f^{X}_{\nu}\xi^{\nu}_{X}+\nabla_{\nu}\left(\frac{1}{2}\sqrt{-g}\mu^{\nu\lambda\tau}_{X}n_{\nu\lambda\tau}^{X}\xi^{\mu}_{X}\right),\nonumber\\
\label{Lag4}
\end{eqnarray}
where $X=n_{(DM)}, s_{(DM)}, n_{(DE)}, s_{(DE)}$; are number and entropy densities for $DM$ and $DE$ respectively.$f^{X}_{\nu}=-(\nabla_{\mu}n^{\mu})\mu^{X}_{\nu}$ is the force density guaranteeing the conservation requirement $\nabla_{\mu}{T^{\mu}}_{\nu}=0.$ This force encodes all external forces and dissipation. The last term in equation (\ref{Lag4}) is a fluid boundary term whose form guarantees the vanishing of  ${\xi^{\mu}}_{X}$ leading to a well-posed action. ${\mu^{X}}_{\nu}$ takes the form in equation (\ref{entr}). The last two terms in equation (\ref{Lag4}) are critical to our understanding of how $DM$ and $DE$ interact. In the case of entrainment, the momentum of one species carries with it mass current of the other species. Entrainment is an observable effect in lab experiments. We see no reason, technology permitting, why this should not be the case in cosmological experiments. In particular, we ask if entrainment can leave an imprint on the {\it Cosmic Microwave Background} and if such is detectable? We will pursue this elsewhere \cite{BTcmb}. As in \cite{And5}, $\mu^{\mu}_{X}$ is the canonical conjugate momentum related to the flux $n^{\mu}_{X}$, from which it is possible to derive a related vorticity parameter; 
\begin{eqnarray}\label{vort} 
2\nabla_{[\mu}\mu^{X}_{\nu]}=\omega^{X}_{\mu\nu}.
\end{eqnarray}   

It is straight forward to include other couplings in the action. For example we could include the variation of $Maxwell$ action $S_{MAX}=1/16\pi \left(\int_{M}d^4 x\sqrt{-g}F_{\mu\nu}F^{\mu\nu}\right)$ and $Coulomb$ action $S_{C}=\left(\int_{M}d^4 x\sqrt{-g}\sum_{X} j^{\nu}_{X}A_{\nu}\right)$.
In particular, varying $Maxwell$ action with respect to the vector potential $A_{\nu}$ and the metric $g_{\nu\lambda}$ and $Coulomb$ action with respect to the flux $n^{\mu}_{X}$, the vector potential $A_{\nu}$ and the metric $g_{\nu\lambda}$ yields
\begin{widetext}
\begin{eqnarray}\label{Act2}
\delta S_{Max}+\delta S_{C}=&\int_{M}&d^{4}x\sqrt{-g}\left[{\sum_{\alpha}(e_{X}A_{\mu})\delta n_{X}^{\mu}}+\frac{1}{4\pi}\left(\nabla_{\nu}F^{\lambda\mu}+4\pi\sum_{X}j^{\mu}_{X}\right)\delta A_{\mu}\right]\nonumber\\
&+\int_{M}&d^{4}x\sqrt{-g}\left[\frac{1}{2}j^{\lambda}_{X}A_{\lambda}g^{\mu\nu}-\frac{1}{32\pi}(F_{\lambda\gamma}F^{\lambda\gamma}g^{\mu\nu}-4 F^{\mu\lambda}F^{\nu}_{\lambda})\right].
\end{eqnarray}
\end{widetext}
The unconstrained variation of action for this minimally coupled fluid mixture is then given by the sum \[\delta S=\delta S_{M_{(X)}}+\delta S_{MAX}+\delta S_{C},\] subject to equation (\ref{Act2}). It was pointed out in \cite{And4} that the minimal coupling considered above leads to a modified conjugate momentum of the form; $\tilde{\mu}^{X}_{\mu}={\mu}^{X}_{\mu}+e^{X}A_{\mu}$ which can be seen when coefficients of $\delta n_{X}^{\mu}$ are collated. Although one can obtain field equations from the variation of these actions, the establishment of the equations demand that the modified momentum vanishes. This vanishing in turn implies that matter or energy must necessarily be absent. This counter-intuitive finding is a manifestation of the unconstrained variation based on $Euler$ type variation (the reader is referred to appendix (\ref{sec4.0}) and section (\ref{convec})). In effect, one would like the modified conjugate momentum $\tilde{\mu}^{\alpha}_{a}$ to have the same status as the unmodified one ${\mu}^{X}_{\mu}$ which is achieved via a $Lagrangian$ variation (see appendix (\ref{pullb})).
 As discussed above, the energy functional $\Lambda$ is a function of co-moving number densities and is, in turn, the carrier of microphysics from the $EoS$. Although we have considered a multi-fluid environment with at least one species of particles being charged, the coupling to electromagnetism is however not encoded in the energy functional \cite{And2}. The matter-spacetime coupling is obtained from varying $\Lambda$ with respect to the metric $g_{\mu\nu}$. This yields the matter stress-energy tensor $T^{\mu\nu}_{M}$.
In particular, keeping the matter fluxes constant and varying the energy functional gives  
\begin{widetext}
\begin{eqnarray}
 T^{\mu\nu}\delta g_{\mu\nu}\equiv\frac{2}{\sqrt{-g}}\delta(\sqrt{-g}\mathcal{L})=\bigg[(\Lambda-\sum_{X}n^{\mu}_{X}\mu^{X}_{\mu}){\delta^{\mu}}_{\lambda}+\sum_{X}{n^{\mu}}_{X}{\mu^{X}}_{\lambda}\bigg]g^{\lambda\nu}\delta g_{\mu\nu}
\end{eqnarray}
\end{widetext}
where again $X=n_{(DM)}, s_{(DM)}, n_{(DE)}, s_{(DE)}$. Coupling to electromagnetism, is expressed in terms of
\begin{eqnarray}
F_{\mu\nu}=2\nabla_{[\mu}A_{\nu]},
\end{eqnarray} 
where $A_{\nu}$ is the electromagnetic vector potential and $F_{\mu\nu}$ is the $Faraday$ anti-symmetric tensor. The coupling is via the matter flow and the charge current $j^{a}$ (the current is conserved in this case: $\nabla^{\mu}j_{\mu}=0$). In this multi-fluid system environment, the current is the sum of the individual currents and is given by $j^{\mu}=\Sigma_{X}e_{X}n_{(X)}^{\mu}$. One easily recovers $Maxwell's$ equations if they vary the electromagnetic term of the $Lagrangian$ while keeping the current fixed. In particular, it is straightforward to show that $\nabla_{\mu}F^{\mu\nu}=\mu_{0}j^{\nu}$ where $\mu$ is the coupling constant. A variation with respect to the metric yields the electromagnetic part of the stress-energy momentum tensor. 
\begin{eqnarray}
T^{EM}
_{\mu\nu}=\frac{1}{\mu_{0}}\left(g^{\lambda\gamma}F_{\mu\lambda}F_{\nu\gamma}-\frac{1}{4}(F_{\lambda\gamma}F^{\lambda\gamma})\right),
\end{eqnarray} 
where $T^{EM}_{\mu\nu}=j_{\mu}F^{\mu\nu}\equiv -f^{\nu}_{L}$. $f^{\nu}_{L}$ is the $Lorentz$ force. This means that a coupling involving one charged species alters equation (\ref{vort}) to
\begin{eqnarray}\label{encod}
2n^{\nu}_{X}\nabla_{[\nu}{\mu}^{X}_{\mu]}+\Gamma_{X}{\mu}^{X}_{\mu}=j^{\nu}_{X}F_{\mu\nu}-\Gamma_{X}e_{X}A_{\mu}+R^{X}_{\mu}.\nonumber\\
\end{eqnarray} 
where $R^{X}_{\mu}$ is the resistivity parameter and encodes dissipation. Since
\begin{eqnarray}
\mu^{X}_{\mu}=g_{\mu\nu}\bigg(\mathcal{B}^{X}{n^{\nu}}_{X} +\sum_{X\ne Y}\mathcal{A}^{XY}n^{\nu}_{Y}\bigg),
\end{eqnarray} 
It is clear that entrainment  given by $\mathcal{A}^{XY}$ contributes to $\mu^{X}_{\mu}$ which in turn is linked to the magnetic potential and resistivity parameter in equation (\ref{encod}). The entrainment effects are embedded in these couplings and hidden in the resulting equations of motions. This suggests, for example, that the dark-sector may indirectly affect the evolution of magnetic fields or analogues thereof \cite{bobAn1, bobAn2, bobAn3, bobAn4}. We will investigate the full impact of this implication elsewhere\cite{BTcmb}. The total energy stress-tensor is therefore, given by
\begin{eqnarray}
T^{\mu\nu}=T_{M}^{\mu\nu}+T_{EM}^{\mu\nu},
\end{eqnarray} 
and it can be shown that $\nabla_{\mu}T^{\mu\nu}=0.$
\section{\label{sec5} Conclusion} 
We have examined (i) single-fluid and multi-fluid approximations of relativistic fluids from the flux point of view, (ii) the convective variational approach, (iii) formulation of convective variational approach for $DM$ interacting with $DE$, and (iv) entrainment effect in this interaction. We find that the various couplings and particularly {\it entrainment} allows for the dark-sector candidates to affect the dynamics of the constituent fluids. Such interactions may be interesting for the solution of the coincidence problem for example. The formalism we have examined here may also be useful in distinguishing cosmological features of these couplings, something that can be probed by current cosmological observations. This would enable us to place constraints on the nature of the interaction considered. It would be interesting to extend this formalism to a general formalism to study the growth of $DM$ perturbations in the presence of interactions between $DM$ and $DE$. This could then allow for the examination of the signature of such interactions on the temperature anisotropies of the large-scale cosmic microwave background (CMB). It has been found \cite{Bruck} that the effect of such interactions has a significant signature on both the growth of $DM$ structure and the late integrated $Sachs$ $Wolfe$ effect($ISW$) given a single-fluid approximation. How would this change, given the multi-fluid approximation examined here? This will be examined in \cite{BTcmb}.

\section{Acknowledgements}
We acknowledge that TO is funded through the $UCT$ postgraduate funding office (South Africa) and BO is funded by $URC$ and $NGP$($UCT$). 
\appendix
\section{\label{convec} Convective Variational Approach}
This section and the next reviews basic formulations of the $Lagrangian$ approach found in \cite{Cart0, Prix1}. The key quantity in the convective variational framework is the master function $\Lambda=\Lambda(n^{\mu}_{X}, g_{\mu\nu})$, where, $n^{\mu}_{X}$ is the number current or flux, $\mu$ is a spacetime index, $X$ is a constituent index, $\nu$ is a spacetime index, and $g_{\mu\nu}$ is the spacetime metric. Related to the number currents are the momenta $\mu^{X}_{\alpha}$. It is easy to show that the standard general variation leads to
\begin{eqnarray}\label{app0}
\delta\Lambda&=&\sum_{X}\mu^{X}_{\alpha}\delta n^{\alpha}_{X}+\frac{\partial\Lambda}{\partial g^{\mu\nu}}\delta g^{\mu\nu}
\end{eqnarray}
where $\mu^{X}_{\alpha}$ are the momenta related to the constituent $X$. We will require that $\delta n^{\alpha}_{X}=0=\delta g^{\mu\nu}$ as dictated by the principle of least action. The conservation of the stress-energy momentum tensor $\nabla^{\mu}T_{\mu\nu}=0$ arises from consideration synonymous with electromagnetism. In particular, beginning with $Euler$ relation, equation (\ref{Eulers}) can be written in the form \[\rho+p=\sum_{X}\mu^{X}n^{X}+TS.\]. One can show that $d\rho=-\mu^{X}_{\nu}dn^{\nu}_{X}$, where we have used the definition $\mu_{\nu}=\mu u_{\nu}$ with the condition $u_{\mu}du^{\mu}=0$. Indeed, from 
\begin{eqnarray}
{T^{\mu}}_{\nu}=p{\delta{\mu}}_{\nu}+\sum_{X}n^{\mu}_{X}\mu^{X}_{\nu}.
\end{eqnarray} 
It can be shown that the variational framework suggests that the equations of motion can be written as a force-balance equation,
\begin{eqnarray}\label{app2}
\nabla^{\mu}T_{\mu\nu}&=&\sum_{X}{f^{X}_{\nu}}=0
\end{eqnarray}
where the generalised force density works out to be
\begin{eqnarray}\label{app3}
f^{X}_{\nu}&=&\mu^{X}_{\nu}\nabla_{\sigma}n^{\sigma}_{X}+n^{\sigma}_{X}\nabla_{[\sigma}\mu^{X}_{\nu]}
\end{eqnarray}
and where $\alpha$, and $\sigma$ are spacetime indices. With the above information, the stress-energy tensor can be derived. From equations (\ref{app2}) and (\ref{app3}), it can be shown that,
\begin{eqnarray}\label{app4}
\nabla_{\mu}T^{\mu}_{\nu}&=&\sum_{X}\bigg[\mu^{X}_{\nu}\nabla_{\mu}n^{\mu}_{X}+n^{\mu}_{X}(\nabla_{\mu}\mu^{X}_{\nu}-\nabla_{\nu}\mu^{X}_{\mu})\bigg]
\end{eqnarray}
since,
\begin{eqnarray}
\nabla_{[\mu}\mu^{X}_{\nu]}&=&\nabla_{\mu}\mu^{X}_{\nu}-\nabla_{\nu}\mu^{X}_{\mu}
\end{eqnarray}
for this case. Equation (\ref{app4}) then becomes,
\begin{eqnarray}\label{app7}
\nabla_{\mu}T^{\mu}_{\nu}&=&\sum_{X}\bigg[\mu^{X}_{\alpha}\nabla_{\nu}n^{\alpha}_{X}-n^{\alpha}_{X}\nabla_{\nu}\mu^{X}_{\alpha}-\mu^{X}_{\alpha}\nabla_{\nu}n^{\alpha}_{X}\nonumber\\&+&\Psi\nabla_{\mu}g^{\mu}_{\nu}
+n^{\mu}_{X}\nabla_{\mu}\mu_{\nu}+\mu_{\nu}\nabla_{\mu}n^{\mu}_{X}\bigg]
\end{eqnarray}
where,
\begin{eqnarray}\label{Psi}
\Psi&=&\Lambda-\sum_{X}\mu^{X}_{\alpha}n^{\alpha}_{X}
\end{eqnarray}
The metric $g_{\alpha\sigma}$ has symmetry condition
\begin{eqnarray}\label{app6}
g_{\alpha\sigma}&=&g_{(\alpha\sigma)}
\end{eqnarray} 
regardless of whether they are physically permitted. With the conditions above, the most general infinitesimal variation of the master function $\Lambda$ that can be envisaged will have the form in equation (\ref{app0}), where $n^{\alpha}_{X}$ is a set of vectors representing diverse currents of entropy and whatever kinds of neutral or charged, not necessarily conserved particles, with the specification of the partial derivatives completed in view of equation (\ref{app6}) by the appropriate symmetry condition
\begin{eqnarray}
\frac{\delta\Lambda}{\delta g_{\alpha\sigma}}&=&\frac{\delta\Lambda}{\delta g_{\sigma\alpha}}
\end{eqnarray}
A variation is given by,
\begin{eqnarray}
\delta\Lambda&=&\xi^{\mu}\nabla_{\mu}\Lambda
\end{eqnarray}
where $\xi^{\mu}$ is an arbitrary infinitesimal displacement vector field. The infinitesimal variations of the variables appearing in equation (\ref{app0}) are given by the corresponding $Lie$ derivatives, namely,
\begin{eqnarray}
\delta n^{\alpha}_{X}&=&\xi^{\nu}\nabla_{\nu}n^{\alpha}_{X}-n_{X}^{\nu}\nabla_{\nu}\xi^{\alpha}
\end{eqnarray}
\begin{eqnarray}
\delta g_{\alpha\sigma}&=&2\nabla_{(\alpha}\xi_{\sigma)}
\end{eqnarray}
With the above information, it can be shown that,
\begin{eqnarray}
\Bigg[\sum_{X}\mu^{X\alpha}n^{\sigma}_{X}-2\frac{\partial\Lambda}{\partial g_{\alpha\sigma}}\Bigg]\nabla_{\nu}\xi_{\mu}&=&\Bigg[\sum_{X}\mu^{X}_{\alpha}\nabla_{\nu}n^{\alpha}_{X}-\nabla_{\nu}\Lambda\Bigg]\xi^{\nu}\nonumber\\
\end{eqnarray}
Since $\xi^{\nu}$ and its covariant derivative at any point are arbitrary, the corresponding coefficients must vanish identically, the resulting $Noether$ identities thus being the obvious relation,
\begin{eqnarray}\label{Lamo}
\nabla_{\nu}\Lambda&=&\sum_{X}\mu^{X}_{\alpha}\nabla_{\nu}n^{\alpha}_{X}
\end{eqnarray}
together with the less trivial relation,
\begin{eqnarray}
2\frac{\partial\Lambda}{\partial g_{\alpha\sigma}}&=&\sum_{X}\mu^{X\alpha}n^{\sigma}_{X}
\end{eqnarray}
This means equation (\ref{app7}) can be written as,
\begin{eqnarray}
\nabla_{\mu}T_{\nu}^{\mu}&=&g^{\mu}_{\nu}\nabla_{\mu}\Psi+\Psi\nabla_{\mu}g^{\mu}_{\nu}+n^{\mu}_{X}\nabla_{\mu}\mu_{\nu}+\mu_{\nu}
\nabla_{\mu}n^{\mu}_{X}\nonumber\\
\end{eqnarray}
after using equations (\ref{Psi}) and (\ref{Lamo}). It follows that
\begin{eqnarray}
T^{\mu}_{\nu}&=&\Psi g^{\mu}_{\nu}+\sum_{X}\mu_{\nu}n^{\mu}_{X}
\end{eqnarray}
If we introduce a set of convection vectors, $\beta^{\mu}_{X}$ such that $
h_{X}\beta^{\mu}_{X}=n^{\mu}_{X}$, and $\mu^{X}_{\nu}\beta^{\nu}_{X}=-1$ then
$h_{X}=-\mu^{X}_{\mu}n^{\mu}_{X}.$ The convection vector can then be used as a projection operator as follows
\begin{eqnarray}
\perp_{X}^{\mu\nu}&=&g^{\mu\nu}+\mu^{\mu}_{X}\beta^{\nu}_{X}
\end{eqnarray}
and hence
\begin{eqnarray}
\perp^{\mu}_{X\nu}\beta^{\nu}_{X}&=&\perp^{\mu\nu}_{X}\mu^{X}_{\nu}=0.
\end{eqnarray}
It can also be shown that,
\begin{eqnarray}
\nabla_{\mu}n^{\mu}_{X}&=&-\beta^{\mu}_{X}f^{X}_{\mu}
\end{eqnarray}
and 
\begin{eqnarray}
\perp^{X}_{\nu\sigma}f^{\sigma}_{X}&=&h_{X}\mathcal{L}_{X}\mu^{X}_{\nu}.
\end{eqnarray}
The convection vector can be decomposed with respect to the 4-velocity $u^{\mu}$ and a drift velocity $v^{\mu}_{X}$ as follows \begin{eqnarray}
\beta^{\mu}_{X}&=&\beta_{X}[u^{\mu}+v^{\mu}_{X}]\label{drift}
\end{eqnarray}
$\mu^{X}=\frac{1}{\beta_{X}}$ represents a chemical type potential for species $X$ with respect to the chosen frame. The relative velocity or drift velocity vector in equation (\ref{drift}) is restricted to satisfy the orthogonality condition,
\begin{eqnarray}
u_{\mu}v^{\mu}_{X}&=&0
\end{eqnarray}
while the four velocity also satisfies the condition below,
\begin{eqnarray}
u_{\mu}u^{\mu}=-1
\end{eqnarray}
The above results, then lead to the suggestion that,
\begin{eqnarray}
u^{\sigma}\nabla_{\alpha}T^{\alpha}_{\sigma}+\sum_{X}[\mu^{X}\nabla_{\sigma}n^{\sigma}_{X}+v^{\sigma}_{X}f^{X}_{\sigma}]&=&0
\end{eqnarray}
To see how things work out in the present formalism, the entropy fluid [with index $S$] is singled out by defining $s^{\mu}=n^{\mu}_{S}$ and $T=\mu_{S}$ (the alternative representation to the variables in equation (\ref{Lam})). To simplify the final expressions it is also useful to assume that the remaining species are governed by conservation laws of the form
\begin{eqnarray}
\nabla_{\mu}n^{\mu}_{X}&=&\Gamma_{X}
\end{eqnarray}
subject to the constraint of total baryon conservation
\begin{eqnarray}
\sum_{X\neq S}\Gamma_{X}&=&0
\end{eqnarray}
Given this, and the fact that the divergence of the stress-energy tensor should vanish, it can be shown that,
\begin{eqnarray}
T\nabla_{\mu}s^{\mu}&=&-\sum_{X\neq S}\mu^{X}\Gamma_{X}-\sum_{X}v^{\mu}_{X}f^{X}_{\mu}
\end{eqnarray}
Finally, the remaining force contribution can be brought by introducing the linear combination
\begin{eqnarray}
\sum_{X}\zeta^{X}v^{\mu}_{X}&=&0
\end{eqnarray}
constrained by
\begin{eqnarray}
\sum_{X}\zeta^{X}&=&1
\end{eqnarray}
Then defining,
\begin{eqnarray}
\overline{f^{X}_{\nu}}&=&f^{X}_{\nu}
\end{eqnarray}
it can be shown that,
\begin{eqnarray}
T\nabla_{\mu}s^{\mu}&=&-\sum_{X}\mu^{X}\Gamma_{X}-v^{\mu}_{X}\overline{f^{X}_{\nu}}\nonumber\\
&\geq&0
\end{eqnarray}
The two terms in this expression represent, respectively, the entropy increase due to $(i)$ chemical reactions and $(ii)$ conductivity. The simplest way to ensure that the second law of thermodynamics is satisfied is to make each of the two terms positive definite. A reasonable starting point would be to assume that each term is linear. For the chemical reactions, this would mean that $\Gamma_{X}$ is expanded according to
\begin{eqnarray}
\Gamma_{X}&=&-\sum_{Y\neq S}\mathcal{C}_{XY}\mu^{Y}
\end{eqnarray}
where $\mathcal{C}_{XY}$ is a positive definite matrix composed of the various reaction rates. Similarly, for the conductivity term it is natural to consider standard resistivity such that
\begin{eqnarray}
\overline{f^{X}_{\alpha}}&=&-\sum_{Y}\mathcal{R}^{XY}_{\alpha\sigma}v^{\sigma}_{Y}.
\end{eqnarray}

\section{\label{pullb}Pull Back Formalism, Three-Forms and Convective Variational Formalism}
Let $n^{X}_{abc}$ be a three-form that is dual to the flux $n^{d}_{X}$ such that
\begin{eqnarray}
n^{X}_{abc}=\epsilon_{dabc}n^{d}_{X}, n^{a}_{X}=\frac{1}{3}\epsilon^{bcda}n^{X}_{bcd}.
\end{eqnarray}
It has been shown in \cite{And5} that if the convention for transforming between the two dual forms is 
\begin{eqnarray}
\epsilon^{bcda}\epsilon_{ebcd}=3!\delta^{a}_{e},
\end{eqnarray} 
then one can use a well defined {\it pullback}; $Z^{A}_{X}$ that pulls $n^{X}_{abc}$ into the matter space where it takes the identity $n^{X}_{ABC}.$ This means
\begin{eqnarray}
n^{X}_{abc}=\frac{\partial Z^{[A}_{X}}{\partial x^{a}}\frac{\partial Z^{B}_{X}}{\partial x^{b}}\frac{\partial Z^{C]}_{X}}{\partial x^{c}}n^{X}_{ABC}.
\end{eqnarray} 
A similar development can be done for the chemical potential $\mu^{abc}_{X}$, where $Z^{A}_{X}$ is now a push-forward. 
It is straight forward to show that $Z^{A}_{X}$ is {\it Lie}-dragged along individual fluid world-lines leading to its conservation. In particular 
\begin{eqnarray}
\frac{dZ^{A}_{X}}{d\tau_{X}}&=&u^{a}_{X}\nabla_{a}X^{A}_{X}=0.
\end{eqnarray}
$Z^{A}_{X}$ is an unconstrained scalar that we can subject the variational principle with the hope of obtaining field equations for the fluxes. It follows that one can use $Lagrangian$ displacement $\xi^{\mu}_{X}$, as pointed out in \cite{And5}, to link variations of matter space variables to spacetime variables. One can define a relativistic $Lagrangian$ variation associated with this displacement \cite{And5}. i.e.
$\Delta_{X}\equiv\delta+\mathcal{L}_{\xi_{X}}$, where the first term is $Euler's$ variation and the second term is the $Lie$derivative. In terms of this variation, it follows that
\begin{eqnarray}
\Delta_{X} Z^{A}_{X}&=&\delta Z^{A}_{X}+\mathcal{L}_{\xi_{X}}Z^{A}_{X}=0.
\end{eqnarray} 
This clearly shows that $\delta Z^{A}_{X}=-(\nabla_{\mu}Z^{A}_{X})\xi^{\mu}_{X},$ from which one can show that $\Delta_{X}n^{X}_{abc}=0.$ Alternative and comparative variational formalisms to the one above is developed in the next section. It suffices to say that we now have relativistic $Lagrangian$ variational formalism that we need for our study.
\section{\label{sec4.0}An  Alternative Formulation of the Convective Formalism: The Prix Method \cite{Prix1}}
Let us consider particle flow lines represented by \[x^{i}=x^{i}({\bf a},t)\] where ${\bf a}^{i}$ are particle coordinates for the individual particles. This `material space for this particle' is related to the `physical space' $x^{i}$ as indicated. Now assume an infinitesimal spatial displacement $\xi^{i}({\bf x},t)$ and temporal shifts $\tau({\bf x},t)$ of the fluid particle flow induced variations of fluid variables given by
\begin{eqnarray}
x'({\bf a},t')&=&x^{i}({\bf a},t)+\xi^{i}({\bf x},t)\\
t'&=&t+\tau ({\bf x},t).
\end{eqnarray} 
Any scalar type physical quantity $Q({\bf x},t)$ is changed by this transformation to $Q({\bf x}', t')$. 

Defining $Euler$ and $Lagrangian$ variations by
\begin{eqnarray}
\delta Q&\equiv&Q'({\bf x},t)-Q({\bf x},t)\\
\Delta Q&\equiv&Q'({\bf a},t')-Q({\bf a},t)= Q'({\bf x}',t)-Q({\bf x},t),\nonumber\\
\end{eqnarray} 
one can show, using $Taylor$ expansion to linear-order, that 
\begin{eqnarray}
\Delta Q=\delta Q+\xi^{j}\nabla_{j}Q({\bf x},t)+\tau\partial_{t}Q({\bf x},t).
\end{eqnarray} 
How does velocity change given these infinitesimal changes?
\begin{eqnarray}
v'({\bf a},t')&=&d_{t'}x({\bf a}, t)+d_{t}\xi({\bf x},t)+\mathcal{O}(2).
\end{eqnarray} 
Therefore, to linear order
\begin{eqnarray}\label{vel1}
v'^{i}({\bf a},t')&=&d_{t}x^{i}({\bf a}, t)\left[1-d_{t}\tau\right]+d_{t}\xi^{i}({\bf x},t)
\nonumber\\&=&v^{i}-v^{i}d_{t}\tau+d_{t}\xi^{i}({\bf x},t),
\end{eqnarray} 
where products involving more than one spatial or temporal derivatives are treated as second order and discarded in this approximation. It is therefore clear from equation (\ref{vel1}) that to linear order
\begin{eqnarray}
\Delta v^{i}&=&\partial_{t}\xi^{i}+v^{l}\nabla_{l}\xi^{i}-v^{i}\partial_{t}\tau-v^{i}v^{l}\nabla_{l}\tau,
\end{eqnarray} 
where $d_{t}=\partial_{t}+v^{l}\nabla_{l}$ 
It is also possible to determine how the $Jacobian$ is affected by these changes. In particular
\begin{eqnarray}
\mathcal{J}'^{i}_{j}
&=&\frac{\partial x^{i}({\bf a},t)}{\partial a_{j}}+\frac{\partial x^{i}({\bf a},t)}{\partial t}\frac{\partial t}{\partial a_{ij}}\vert_{t'} +\frac{\partial \xi^{i}({\bf a},t)}{\partial a_{j}}\nonumber\\
\mathcal{J}'^{i}_{j}({\bf a},t')&=&\mathcal{J}^{i}_{j}({\bf a},t)+\nabla_{j}\xi^{j}-v^{i}\nabla_{j}\tau\label{prix2},
\end{eqnarray} 
wherein we have used $Taylor$ expansion to linear order. The $Lagrangian$ variation to the $Jacobian$ may be written as 
\begin{eqnarray}
\Delta\mathcal{J}^{i}_{j}&=&\mathcal{J}^{l}_{j}(\nabla_{l}\xi^{i}-v^{i}\nabla_{l}\tau),
\end{eqnarray} 
where we have used the last line of equation (\ref{prix2}).

Following \cite{Prix1} and using the relationship
\begin{eqnarray}
\frac{\partial det(\bf{A})}{\partial A_{ij}}=det({\bf A})(A^{-1}_{ij})^{T},~~det({\bf A})=|A|
\end{eqnarray} 
it is straightforward to show
\begin{eqnarray}
\frac{\Delta det(\bf{J})}{det({\bf J})}=\nabla_{l}\xi^{l}-v^{l}\nabla_{l}\tau, ~~det({\bf J})=|J|.
\end{eqnarray} 
We can use this relation to monitor how the changes induce variation in the density of $n$ i.e.
\begin{eqnarray}
\Delta n=n\nabla_{l}\xi^{l}-nv^{l}\nabla_{l}\tau.
\end{eqnarray} 
It is now possible to combine density and velocity variations to determine the variation in the current associated with a particular species. In particular, for the current $n^{i}=nv^{i}$ we obtain the variation
\begin{eqnarray}
\Delta n^{i}=\left[n\partial_{t}\xi^{i}+n^{l}\nabla_{l}\xi^{i}-n^{i}\nabla_{l}\xi^{l}\right]-n^{i}\partial_{t}\tau.
\end{eqnarray}
It is also possible to incorporate variation of the metric and monitor how such a variation affects the variation of the current. In particular,
\begin{eqnarray}
\Delta n^{i}=\left[n\partial_{t}\xi^{i}+n^{l}\nabla_{l}\xi^{i}-n^{i}\nabla_{l}\xi^{l}\right]-n^{i}\partial_{t}\tau-\frac{1}{2}n^{i}g^{lj}\delta g_{lj}\nonumber\\
\end{eqnarray}

\end{document}